\documentclass[aps,prd,nofootinbib,superscriptaddress,preprint]{revtex4}
\usepackage{amssymb}
\usepackage{amsmath,color}
\usepackage{epsfig}
\usepackage{graphicx,epsf,epsfig}
\usepackage{bbm}
\usepackage{subfigure}
\usepackage{multirow}
\usepackage{setspace}
\usepackage{verbatim}

\newcommand{\ba}{\begin{array}{c}}
\newcommand{\baz}{\begin{array}{cc}}
\newcommand{\bad}{\begin{array}{ccc}}
\newcommand{\bav}{\begin{array}{cccc}}
\newcommand{\baf}{\begin{array}{ccccc}}
\newcommand{\ea}{\end{array}}

\def\be{\begin{equation}}
\def\ee{\end{equation}}
\def\gs{\mathrel{
   \rlap{\raise 0.511ex \hbox{$>$}}{\lower 0.511ex \hbox{$\sim$}}}}
\def\ls{\mathrel{
   \rlap{\raise 0.511ex \hbox{$<$}}{\lower 0.511ex \hbox{$\sim$}}}}

\newcommand{\bea}{\begin{equation} \begin{array}{c}}
\newcommand{\eea}{ \end{array} \end{equation}}

\def\slc#1{\setbox0=\hbox{$#1$}           
    \dimen0=\wd0                                 
    \setbox1=\hbox{/} \dimen1=\wd1               
    \ifdim\dimen0>\dimen1                        
       \rlap{\hbox to \dimen0{\hfil/\hfil}}      
       #1                                        
    \else                                        
       \rlap{\hbox to \dimen1{\hfil$#1$\hfil}}   
       /                                         
    \fi}
\begin{document}
\title{RG running in a minimal UED model in light of recent LHC Higgs mass bounds}
\date{\today}

\author{Mattias Blennow}
\email{Mattias.Blennow@mpi-hd.mpg.de}

\affiliation{Max-Planck-Institut f{\"u}r Kernphysik, Postfach
103980, 69029 Heidelberg, Germany}

\author{Henrik Melb{\'e}us}
\email{melbeus@kth.se}

\author{Tommy Ohlsson}
\email{tommy@theophys.kth.se}

\affiliation{Department of Theoretical Physics, School of
Engineering Sciences, KTH Royal Institute of Technology --
AlbaNova University Center, Roslagstullsbacken 21, 106 91 Stockholm,
Sweden}

\author{He Zhang}
\email{he.zhang@mpi-hd.mpg.de}

\affiliation{Max-Planck-Institut f{\"u}r Kernphysik, Postfach
103980, 69029 Heidelberg, Germany}

\begin{abstract}
We study how the recent ATLAS and CMS Higgs mass bounds affect the renormalization group running of the physical parameters in universal extra dimensions. Using the running of the Higgs self-coupling constant, we derive bounds on the cutoff scale of the extra-dimensional theory itself. We show that the running of physical parameters, such as the fermion masses and the CKM mixing matrix, is significantly restricted by these bounds. In particular, we find that the running of the gauge couplings cannot be sufficient to allow gauge unification at the cutoff scale.
\end{abstract}

\maketitle


\section{Introduction}

Recently, the ATLAS and CMS collaborations presented new bounds on the mass of the Standard Model (SM) Higgs boson, excluding values outside the range 115.5~GeV - 131~GeV \cite{Atlas:2011} and 115~GeV - 127~GeV \cite{CMS:2011}, respectively, at 95~\% confidence level. While the search for the Higgs boson is the primary goal of the LHC, the experimental collaborations are also intensively searching for signs of new physics beyond the SM. Among the most popular models describing new physics within the reach of the LHC is the universal extra dimensions (UED) model \cite{Appelquist:2000nn}. In this model, all of the SM fields are promoted to a higher-dimensional spacetime, giving rise to infinite Kaluza--Klein (KK) towers. The lowest-mass KK modes are usually assumed to be located at the TeV scale, and in particular, the lightest KK particle could be an interesting dark matter candidate \cite{Servant:2002aq,Cheng:2002ej}.

An important feature of extra-dimensional models is the impact of the large number of KK modes on the renormalization group (RG) running of physical parameters. The RG running in extra-dimensional models has previously been investigated, e.g., in Refs.~\cite{Dienes:1998vh,Dienes:1998vg,Abel:1998wa,Bhattacharyya:2002nc,Bhattacharyya:2006ym,Cornell:2010sz,Cornell:2011ge,Blennow:2011mp}. It has been shown that the RG evolution changes from the typical logarithmic running in four-dimensional models to an effective power-law running at high energies. This means that sizable running could take place at relatively low energy scales. In particular, the possibility of achieving gauge coupling unification at intermediate energy scales has been discussed \cite{Dienes:1998vh,Dienes:1998vg}.

In this paper, we use the RG evolution of the Higgs self-coupling constant in order to derive bounds on the minimal five-dimensional UED model, using the recent LHC Higgs mass bounds. Previously, results from LHC Higgs searches have been used to constrain five- and six-dimensional UED models in Refs.~\cite{Nishiwaki:2011gk,Kakuda:2012px}, giving the bound $R^{-1} > 700$~GeV for the minimal five-dimensional UED model. This corresponds to an upper bound $m_H < 500$~GeV on the Higgs boson mass from electroweak precision data \cite{Baak:2011ze}. The RG running of the Higgs self-coupling constant has also recently been used to constrain new physics models in Refs.~\cite{Holthausen:2011aa,Masina:2011aa,EliasMiro:2011aa}.

In addition, we discuss the running of fermion masses and mixing parameters in the UED model, taking the new bounds into account. These fundamental physical parameters are crucial for building new physics models, as well as testing the feasibility of theories beyond the SM. In fact, since the values of these parameters are not predicted by the SM, new physics, which is usually located at some very high energy scale, is needed in order to gain insight into their origin. Thus, we provide values for the fermion masses and mixing parameters at the cutoff scale of the UED model.

The rest of this work is organized as follows: In Sec.~\ref{sec:RGE-ED}, we discuss general features of renormalization group running in extra-dimensional theories. Next, in Sec.~\ref{sec:Higgs}, we discuss the running of the Higgs self-coupling constant and the resulting bounds on the UED model from the LHC Higgs mass bounds. Then, in Sec.~\ref{sec:Gauge}, we show the running of the gauge coupling constants and demonstrate that gauge unification cannot be achieved within the UED model. In Secs.~\ref{sec:FermionMasses} and \ref{sec:CKM}, we give the RG evolution of the fermion masses and the CKM matrix parameters, respectively. Finally, in Sec.~\ref{sec:Summary}, we summarize and discuss our results. In addition, in Appendix \ref{sec:appendix}, we provide the one-loop beta functions that are relevant for our work.


\section{Renormalization group running in extra dimensions}\label{sec:RGE-ED}

A general feature of quantum field theories with extra spatial dimensions is that they are non-renormalizable. However, as pointed out in Ref.~\cite{Dienes:1998vg}, such models could preserve renormalizability if they are truncated at a certain energy scale (i.e., the number of KK modes is finite). In such a situation, physical quantities are subject to a power-law running behavior, in contrast to the typical logarithmic running in ordinary four-dimensional theories. This power-law running may significantly change the running physical parameters, such as the gauge couplings, the quark and lepton masses and mixing angles, and the Higgs mass.

In general, the beta function of a parameter $P$ in an extra-dimensional model can be expressed as
\begin{eqnarray}
    16 \pi^2 \frac{{\rm d}P}{{\rm d}\ln \mu}= \beta_P + s \tilde \beta_P \, ,
\end{eqnarray}
where $\beta_P$ denotes the SM beta function, while $\tilde \beta_P$ corresponds to the contributions from the KK modes at any single KK level to the total beta function. Here, it is assumed that the particle content at each non-zero KK level is the same, except for the particle masses. This is the case in many models, and in particular in the UED model. The scale parameter $s$ is defined as $s = \lfloor \mu / \mu_0 \rfloor$, where $\lfloor x \rfloor$ is the largest integer smaller than $x$, and $\mu_0 \equiv R^{-1}$ is the inverse radius of the extra dimensions, i.e., $s$ counts the number of KK levels below the energy scale $\mu$. At energy scales below $\mu_0$, i.e., below the mass of the lowest KK excitations, the $\tilde \beta_P$ term can be ignored, whereas for $\mu \gg \mu_0$, many KK modes are excited and their contributions change the scale-dependence of the physical parameters from logarithmic to power-law. The relevant one-loop beta functions for the five-dimensional UED model can be found in Appendix~\ref{sec:appendix}.

As mentioned above, in order to make the theory renormalizable, an explicit cutoff scale $\Lambda$ has to be introduced. From this point of view, the UED model is an effective description at low-energy scales, which is replaced by a renormalizable theory above the cutoff scale. In the UED model, $\Lambda$ is usually taken to be the energy scale where the gauge couplings become non-perturbative \cite{Hooper:2007qk}, but it could also be related to a unification scale for the gauge couplings \cite{Dienes:1998vh,Dienes:1998vg}. In this work, we will apply the LHC bounds on the Higgs mass to the running behavior of the Higgs self-coupling constant in order to put bounds on $\Lambda$.

In the following sections, we perform a numerical analysis of the running physical parameters. In our computations, we make use of the full one-loop RGEs without any further approximations. The input values for the physical parameters, at the energy scale $M_Z = 91.2~{\rm GeV}$, are taken from Ref.~\cite{Xing:2011aa}.


\section{Running of the Higgs self-coupling constant}\label{sec:Higgs}

The running of the Higgs self-coupling constant $\lambda$ in the UED
model is given in Eq.~\eqref{eq:RGE-lambda}. Now, the RG evolution
of $\lambda$ can be used to constrain the UED model. In particular,
for different initial conditions, $\lambda$ may approach the
triviality limit ($\lambda$ diverges), or the vacuum stability limit
($\lambda$ becomes negative, i.e., the Higgs potential becomes
unstable). Note that a negative $\lambda$ does not
necessarily mean that the model is invalid, since the electroweak
vacuum might be metastable. In other words, we could live in an
unstable vacuum, while the lifetime of the electroweak vacuum is
longer than the age of the Universe. However, the metastability
limit relies on the fastest process conceived for the transition to
the true vacuum, and any faster process occurring once anywhere in
the Universe would reduce or eliminate the metastability region.
For a general discussion on this topic, see, e.g.,
Ref.~\cite{Sher:1988mj}. Using the relation $\lambda = m_H^2 / v^2$,
this allows us to constrain the parameter space of the UED model
using the LHC Higgs mass bounds, by requiring that neither the
triviality nor the vacuum stability limit is reached below the
cutoff scale. In the low Higgs mass region allowed by the new LHC
bounds, only the vacuum stability condition is important for the UED
model.

\begin{figure*}[ht]
\begin{center}
\includegraphics[width=.75\textwidth]{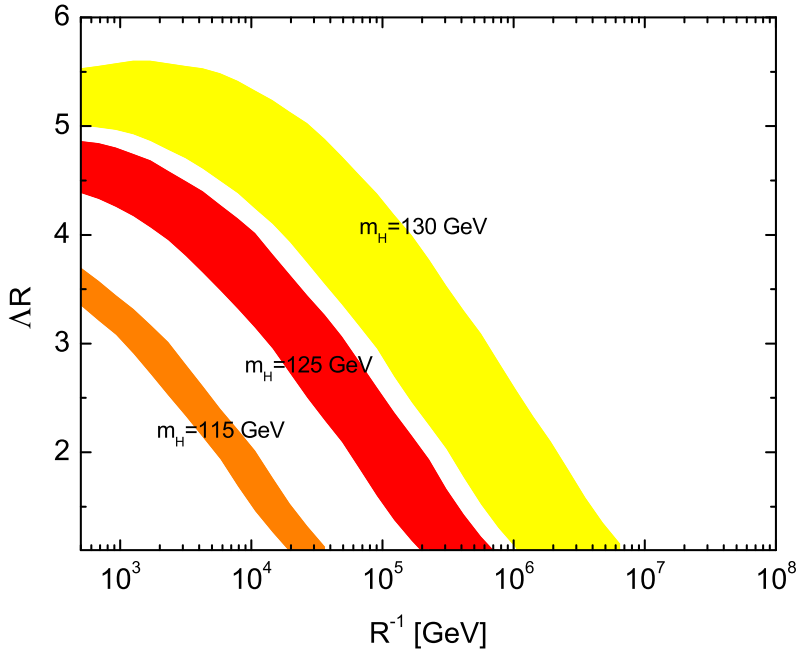}
\caption{\label{fig:lambda} Upper bounds on $\Lambda R$ as a
function of $R^{-1}$ from the vacuum stability condition for the
Higgs self-coupling constant. The bands show the variation of the
bound with the top mass in the range 170.9~GeV - 173.3~GeV, where
the strongest bounds are obtained for the largest value for $m_t$.}
\end{center}
\end{figure*}

In Fig.~\ref{fig:lambda}, we present the bounds on the cutoff scale
$\Lambda$ from the requirement of vacuum stability. We show upper
bounds on the product $\Lambda R$, which counts the number of KK
levels below the cutoff scale, as a function of $R^{-1}$. The
results depend on the value of the top quark mass $m_t$, which is
only known to an accuracy of a few GeV, and therefore, we present
our results for $m_t$ in the range 170.9~GeV - 173.3~GeV. The
weakest bounds are obtained in the phenomenologically interesting
range around 1~TeV. We observe that the global upper limit on the
number of KK modes in the model is five only, constraining the
validity range of the extra-dimensional description significantly.
As we have mentioned, the metastability of the SM vacuum
could result in a smaller lower bound on the Higgs mass, which is the
situation in the UED model. In a semi-classical approximation for
$\lambda \gtrsim -0.05$, the SM Higgs vacuum may still be stable
against the age of the Universe~\cite{Isidori:2001bm}, and thus not
in conflict with experimental observations. This will actually
increase slightly the upper bound on $\Lambda R$ in
Fig.~\ref{fig:lambda}. See Ref.~\cite{EliasMiro:2011aa}
for a detailed numerical analysis on this issue. It should also be noted that our
results rely on the one-loop beta functions, and would be slightly
changed by taking higher-order contributions into account.
Nevertheless, we expect our main conclusions to remain valid at
higher order.


\section{Running of the gauge couplings}\label{sec:Gauge}

Next, the running of the gauge couplings in the UED model is given in Eq.~\eqref{eq:RGE-g}. Solving this equation, we obtain
\begin{eqnarray}\label{eq:RGE-g-analytical}
    \frac{1}{g^2_i(\mu)} = \frac{1}{g^2_i(M_Z)} -\frac{b_i}{8\pi^2} \ln \left(\frac{\mu}{M_Z}\right) - \frac{\tilde b_i}{8\pi^2} \left[ s \ln \left(\frac{\mu}{\mu_0}\right) -\ln s! \right] \; ,
\end{eqnarray}
between the $n$-th and $(n+1)$-th thresholds. The second term in Eq.~\eqref{eq:RGE-g-analytical} corresponds to the SM contributions and the last term to the corrections from the KK modes. Note that the expression in the last parenthesis is always positive, e.g., for $\mu/\mu_0=5,10,40$ one has $s \ln ( \mu / \mu_0 ) - \ln s! \approx 3,8,37$. In the limit of large $s$, i.e., $\mu \gg \mu_0$, the evolution of $g_i (\mu)$ is dominated by the contributions from KK excitations. Since the coefficients $b_i$ and $\tilde b_i$ are in general not the same, the impact on the RG running from the KK modes is different from that from the SM particles. In particular, the sign of $\tilde b_2$ is opposite to that of $b_2$, indicating that $g_2$ tends to increase at higher energy scales. In addition, $\tilde b_1 =27/2$ is larger than the other two coefficients, which leads to a fast running behavior of $g_1$ at high energies.

\begin{figure*}[ht]
\begin{center}
\includegraphics[width=.75\textwidth]{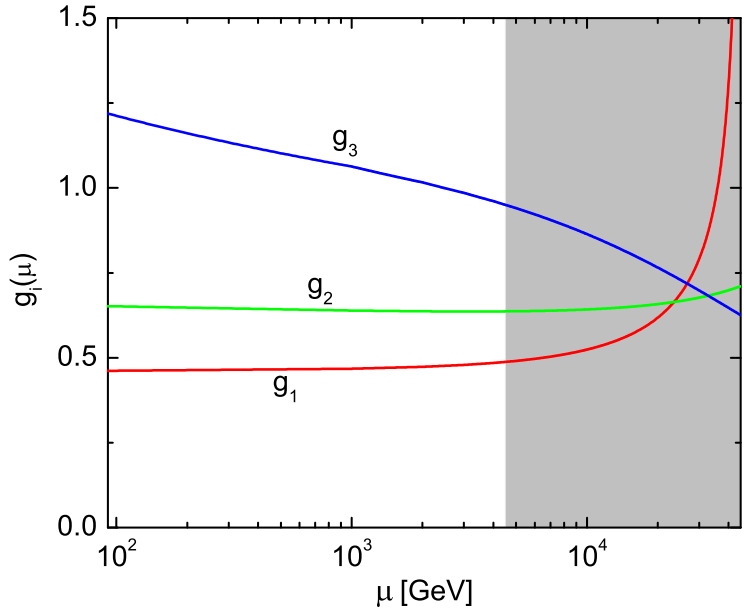}
\caption{\label{fig:fig1} The RG evolution of the three gauge couplings as functions of the energy scale in the UED model, with $R^{-1} = 1~{\rm TeV}$. The gray-shaded area is ruled out by the vacuum stability criterion.}
\end{center}
\end{figure*}

\begin{figure*}[ht]
\begin{center}
\includegraphics[width=.75\textwidth]{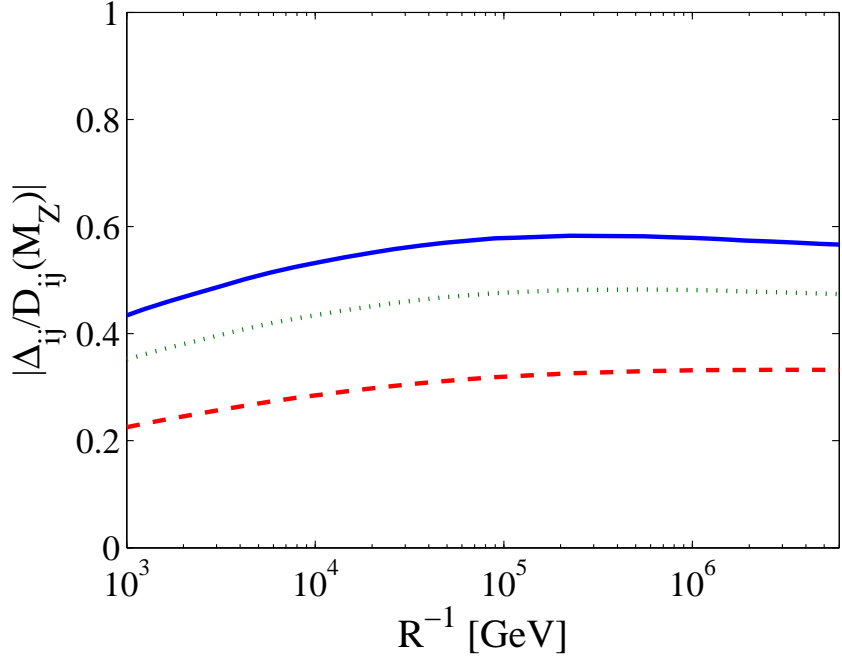}
\caption{\label{fig:nogo} The ratio $\Delta_{ij}(\Lambda)/D_{ij}(M_Z)$, where $\Lambda$ is given by the global upper limit from Fig.~\ref{fig:lambda}. The solid, dashed, and dotted curves correspond to $\{i,j\} = \{1,2\}, \{1,3\}$, and $\{2,3\}$, respectively. A ratio of one or higher would indicate that the gauge couplings $g_i$ and $g_j$ unify below $\Lambda$, which this figure shows is impossible given the current bounds on the Higgs mass.}
\end{center}
\end{figure*}

By solving the RGEs for the gauge couplings, we obtain the running of the $g_i$, which is shown in Fig.~\ref{fig:fig1} for $R^{-1} = 1~{\rm TeV}$. The most interesting feature of this result is that the region where the coupling constants would approximately unify is ruled out by the vacuum stability criterion. In fact, it turns out that this is a general statement and there is a no-go scenario for gauge unification below the cutoff scale $\Lambda$ in the UED model. From Eq.~\eqref{eq:RGE-g-analytical}, it follows that
\begin{equation}
    D_{ij}(\mu) = D_{ij}(M_Z) - \frac{1}{8\pi^2} \left\{ b_{ij}\ln\left(\frac \mu{M_Z}\right) + \tilde b_{ij} \left[s \ln\left(\frac \mu{\mu_0}\right) - \ln s! \right] \right\} \equiv D_{ij}(M_Z) - \Delta_{ij}(\mu),
\end{equation}
where $D_{ij} = 1/g_i^2 - 1/g_j^2$, $b_{ij} = b_i-b_j$, and $\tilde b_{ij} = \tilde b_i - \tilde b_j$. By comparing $\Delta_{ij}(\Lambda)$ with $D_{ij}(M_Z)$, we can observe from Fig.~\ref{fig:nogo} that the ratio never reaches (or is even close to) one. In this figure, we have taken the upper limit of the cutoff scale from Fig.~\ref{fig:lambda} for $m_H = 130$~GeV, which is equivalent to using the global upper limit on $\Lambda$. Thus, $D_{ij}(\mu)$ can never become zero below the cutoff scale, meaning that the gauge couplings $g_i$ and $g_j$ will not unify while the extra-dimensional theory is valid. Note that, while higher-order corrections or allowing for a broader uncertainty range in the input parameters could change the actual values slightly, the ratio is quite far away from one and this conclusion should therefore be robust to such details.


\section{Running of the fermion masses}\label{sec:FermionMasses}

As mentioned before, the evolution of the fermion masses is important for constructing new physics models. Therefore, in this section, we give a detailed discussion on the running fermion masses at the cutoff scale for various radii $R$.

The running quark masses can be obtained from Eq.~\eqref{eq:RGE-Y-quark} as
\begin{eqnarray}
    \dot{m}_f = \left[{\rm Re}(F_u)_{ff} +\alpha_u + s \tilde \alpha_u \right] m_f \; ,
\end{eqnarray}
where $\dot{m}_f \equiv 16\pi^2 {\rm d}m_f / {\rm d} \ln \mu$, $f = u,c,t$, and $F_u$ is given by
\begin{eqnarray}
    F_u & = &  \frac{3}{2} ( D_u - V D_d V^\dagger ) (1+s) \; .
\end{eqnarray}
For the down-type quarks, a similar relation holds, with
\begin{eqnarray}
    F_d & = &  \frac{3}{2} ( V^\dagger D_u V - D_d ) (1+s) \; .
\end{eqnarray}
Here, $V$ denotes the Cabibbo--Kobayashi--Maskawa (CKM) matrix, and
\begin{eqnarray}
    D_u & = &{\rm diag} (y^2_u,y^2_c,y^2_t) \; , \\
    D_d & = & {\rm diag} (y^2_d,y^2_s,y^2_b) \; ,
\end{eqnarray}
where the $y^2_f$ are the eigenvalues of the matrix $Y^\dagger_f Y_f$. We adopt the standard parametrization of the CKM matrix, in which $V$ is parametrized by three mixing angles and one CP-violating phase, viz.,
\begin{eqnarray}\label{eq:para}
    V & = & \left( \begin{matrix}c_{12} c_{13} & s_{12} c_{13} & s_{13} e^{-{\rm i}\delta} \cr -s_{12} c_{23}-c_{12} s_{23} s_{13} e^{{\rm i} \delta} & c_{12} c_{23}-s_{12} s_{23} s_{13} e^{{\rm i} \delta} & s_{23} c_{13} \cr s_{12} s_{23}-c_{12} c_{23} s_{13} e^{{\rm i} \delta} & -c_{12} s_{23}-s_{12} c_{23} s_{13} e^{{\rm i} \delta} & c_{23} c_{13} \end{matrix} \right) \; ,
\end{eqnarray}
with $c_{ij} \equiv \cos \theta_{ij}$ and $s_{ij} \equiv \sin \theta_{ij}$ ($\{i,j\}=\{1,2\}$, $\{1,3\}$, $\{2,3\}$). Note that, above the electroweak symmetry breaking scale, the unbroken gauge symmetry forbids quark and lepton masses. The actual meaning of a fermion mass $m_f$ in this energy region is a measure of the corresponding non-trivial Yukawa coupling eigenvalue $y_f$. We adopt the definition $m_f=y_f v$, above the electroweak scale, where $v \approx 174~{\rm GeV}$ is the vacuum expectation value of the Higgs field in the SM.

In view of the hierarchical spectrum of the quark masses, i.e., $m_t \gg m_{b,c} \gg m_{u,d,s}$, one can neglect all of the Yukawa couplings except for $y_t$, and in this approximation, the running of the quark masses is given by the equations
\begin{eqnarray}\label{eq:mqu}
    \dot{m}_u & \simeq & \left(\alpha_u + s \tilde \alpha_u \right) m_u \; , \\
    \dot{m}_c & \simeq & \left(\alpha_u + s \tilde \alpha_u \right) m_c \; , \\
    \dot{m}_t & \simeq & \left[\frac{3}{2} \left(y_t^2 + s y_t^2 \right) + \alpha_u + s \tilde \alpha_u \right] m_t \; ,
\end{eqnarray}
and
\begin{eqnarray}\label{eq:mqd}
    \dot{m}_d & \simeq & \left[\frac{3}{2}\left(s^2_{13} c^2_{12} c^2_{23}+s^2_{12} s^2_{23} -2s_{12}s_{13}s_{23}c_{12}c_{23} c_\delta\right)\left(y_t^2 + s y_t^2 \right) + \alpha_d + s \tilde \alpha_d \right] m_d \; , \\ \label{eq:mqs}
    \dot{m}_s & \simeq & \left[\frac{3}{2}\left(s^2_{12} s^2_{13} c^2_{23}+s^2_{23} c^2_{12} +2s_{12}s_{13}s_{23}c_{12}c_{23} c_\delta\right)\left(y_t^2 + s y_t^2 \right) + \alpha_d + s \tilde \alpha_d \right]  m_s \; , \\
    \dot{m}_b & \simeq & \left[\frac{3}{2}c^2_{23} c^2_{13}\left(y_t^2 + s y_t^2 \right)+\alpha_d +s \tilde \alpha_d \right] m_b \; .
\end{eqnarray}
The RG evolution equations for $m_u$ and $m_c$ are similar to each other and are both governed by the flavor-diagonal part $\alpha_f$, whereas for the top quark mass, contributions from $y_t$ should be taken into account, which slightly changes the running of $m_t$. In the down-type quark sector, the two light quarks $d$ and $s$ also receive similar RG corrections, since the flavor non-trivial parts [i.e., the $y_t$ terms in Eqs.~\eqref{eq:mqd} and \eqref{eq:mqs}] are suppressed by the CKM mixing angles. Since the $\alpha_f$ parameters are negative, we expect the quark masses to decrease with the energy scale.

As for the charged leptons, we can safely ignore the Yukawa corrections due to the smallness of their masses, and we obtain
\begin{eqnarray}\label{eq:mql}
    \dot{m}_i = \left(\alpha_\ell +s \tilde \alpha_\ell \right) m_i \; ,
\end{eqnarray}
for $i=e,\mu,\tau$, where $\alpha_\ell$ is the flavor-diagonal part of the right-hand side of Eq.~\eqref{eq:beta-l}. Therefore, the charged-lepton masses are essentially rescaled by a common factor at high-energy scales. However, this factor is larger than in the SM, due to the scale parameter $s$. Furthermore, in contrast to the quark sector, the flavor-diagonal part $\alpha_\ell$ is positive, due to the lack of $g_3$ corrections (i.e., leptons do not participate in the strong interactions), which leads to larger values for the charged-lepton masses at higher energies.

\begin{figure*}[ht]
\begin{center}
\includegraphics[width=.75\textwidth]{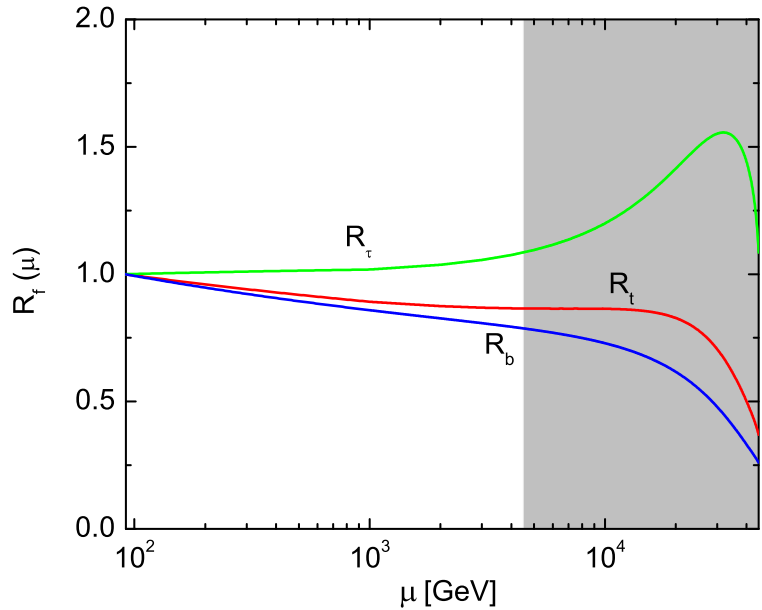}
\caption{\label{fig:m_f} The RG evolution of the mass ratios $R_f$ for the top quark, the bottom quark, and the tau lepton in the UED model, for $R^{-1} = 1$~TeV and $m_H = 125$~GeV. The gray-shaded area is ruled out by the vacuum stability criterion.} \label{fig:Rf}
\end{center}
\end{figure*}

In order to numerically show the RG evolution of the fermion masses, we define the ratios $R_f \equiv m_f(\mu)/m_f(M_Z)$ reflecting the RG corrections at the scale $\mu$. The scale-dependence of $m_t$, $m_b$, and $m_\tau$ is illustrated in Fig.~\ref{fig:Rf}, for $R^{-1}=1~{\rm TeV}$ and $m_H=125~{\rm GeV}$. As expected, the running quark masses decrease with the energy scale. We observe that the quark masses at the cutoff scale are reduced by about 15~\% - 20~\% (c.f., Table~\ref{table:table-I} for detailed numbers). Therefore, it seems impossible to achieve a reasonably good unification of Yukawa couplings, due to the bounds on $\Lambda$. As for the lepton sector, the charged-lepton masses run to larger values at the cutoff scale $\Lambda$, e.g., a 10~\% increase of the tau mass can be observed from the plot. It is also interesting to point out that the values for the running charged-lepton masses~\cite{Xing:2007fb} are maximal at $\mu = 30$~TeV - 40~TeV (or equivalently $n = 30$ - 40), a region which is not allowed by the new Higgs mass bounds. It should be stressed that the running of the charged-fermion masses is not very sensitive to the specific value for the Higgs mass, since $\lambda$ does not enter the beta functions for $Y_f$ at one-loop level. As a reference for model building, we list in Table~\ref{table:table-I} the running quark and charged-lepton masses at the cutoff scale $\Lambda$ for $R^{-1}$ = 500~GeV, 1~TeV, and 10~TeV. These numerical values are consistent with Fig.~\ref{fig:Rf}, and in general, the evolution of the masses is relatively small due to the strong constraints on $\Lambda$. We hope that the table could be useful for building possible extra-dimensional models within the UED framework.

\begin{table}[ht] \vspace{-0.15cm}
\vspace{0.2cm}
\begin{tabular}{c|c|c|c|c}
\hline \hline ~~~~~~~~~~~~~~~~~~~~ & ~~~\multirow{2}{*}{$\mu=M_Z$}
~~~& ~~~$R^{-1} =500~ {\rm TeV}$~~~ & ~~~$R^{-1} =1~ {\rm TeV}$~~~
& ~~~$R^{-1} = 10~ {\rm TeV}$ ~~~\\
& & $\Lambda=2.3~ {\rm TeV}$& $\Lambda=4.5~ {\rm TeV}$&
$\Lambda=36~{\rm TeV}$  \\
\hline  $m_t ~[{\rm GeV}]$&  172 &  160 & 150 & 140 \\
\hline $m_b~[{\rm GeV}]$& $2.86$ & 2.4
& 2.3 & 2.0  \\
\hline $m_c ~[{\rm GeV}]$& $0.638$ & 0.55 & 0.53 & 0.47 \\
\hline $m_s ~[{\rm MeV}]$& $57$ & 49
& 47 & 43 \\
\hline $m_d ~[{\rm MeV}]$& $2.82$ & 2.4
& 2.3 &  2.1 \\
\hline $m_u ~[{\rm MeV}]$& $1.38$ & 1.2 & 1.1 & 1.0 \\
\hline $m_\tau ~[{\rm GeV}]$ & $1.746$ & 1.9
& 1.9 & 1.9  \\
\hline $m_\mu ~[{\rm MeV}]$& $102.7$ & 110 & 110 & 110 \\
\hline $m_e ~[{\rm MeV}]$& $0.4866$ & 0.53 & 0.53 &  0.52 \\
\hline \hline
\end{tabular}\vspace{0.5cm}
\caption{Fermion masses at the cutoff scale $\Lambda$ for $R^{-1}=500~{\rm GeV}$, $R^{-1}=1~{\rm TeV}$, and $R^{-1}=10~{\rm TeV}$, respectively. The input values for the fermion masses at the energy scale $\mu=M_Z$ are listed in the left column for reference.
\label{table:table-I} }
\end{table}

\begin{figure*}[ht]
\begin{center}
\includegraphics[width=.75\textwidth]{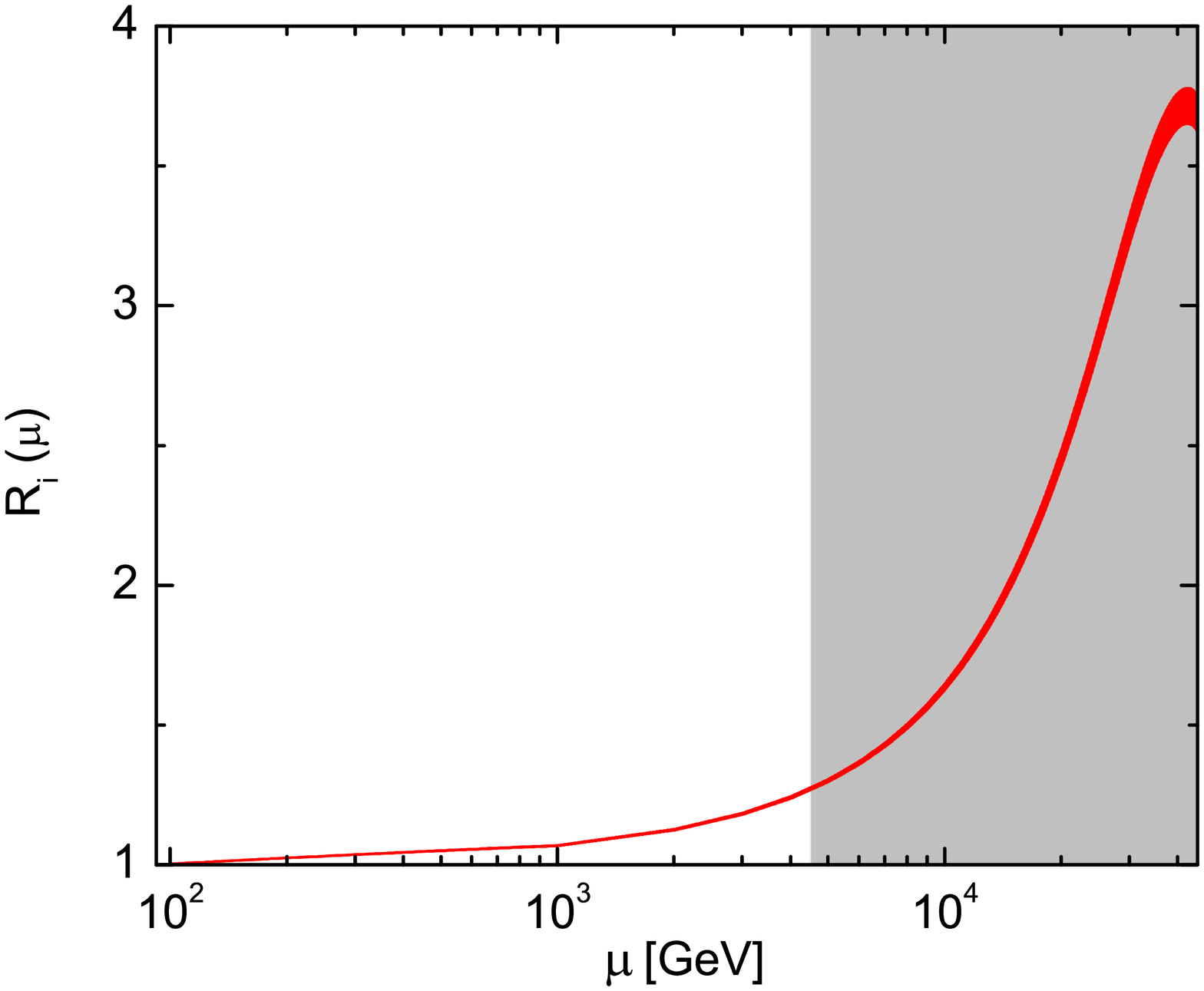}
\caption{\label{fig:Ri} The RG evolution of the neutrino mass ratios $R_i$ in the UED model, for $m_H$ in the range 115~GeV - 130~GeV. The gray-shaded area is ruled out by the vacuum stability criterion.}
\end{center}
\end{figure*}

Finally, we investigate the running neutrino masses with the new Higgs mass bounds in the UED model. By analytically diagonalizing the beta function for $\kappa$, which is given in Eq.~\eqref{eq:RGE-kappa}, we arrive at very compact expressions for the evolution of the neutrino masses
\begin{eqnarray}\label{eq:neutrino-masses}
    \dot m_i & \simeq &  (\alpha_\kappa + s \tilde \alpha_{\kappa})
m_i\, ,
\end{eqnarray}
where $\alpha_\kappa$ is the diagonal part of the right-hand side of Eq.~\eqref{eq:beta-kappa-sm} and we have omitted the charged-lepton Yukawa couplings. An approximate solution to this equation is given by
\begin{eqnarray}\label{eq:mass-approx}
    \frac{m_i(\Lambda)}{m_i (M_Z)} \simeq \left(\frac{\Lambda}{M_Z}\right)^{\alpha_\kappa + s \tilde \alpha_{\kappa}} \, .
\end{eqnarray}
Therefore, the RG running of the neutrino masses is only sensitive to $\alpha_\kappa$, independently of the neutrino mass spectrum and the mixing parameters. Similarly to the charged fermions, we define the ratios $R_i \equiv m_i(\mu)/m_i(M_Z)$ for the neutrino masses, and
illustrate the evolution of $R_i$ in Fig.~\ref{fig:Ri} for $m_H$ in the range 115~GeV - 130~GeV. An important feature of the running, which can be seen from the plot, is that, due to the stability bounds, $R_i$ cannot reach large values below $\Lambda$. Furthermore, the running of the neutrino masses does indeed depend on the Higgs mass $m_H$, since the effective neutrino coupling matrix $\kappa$ receives one-loop corrections from the quartic Higgs interaction.

\section{Running of the CKM mixing matrix}\label{sec:CKM}

The CKM mixing matrix stems from the mismatch between the diagonalization of the Yukawa matrices $Y_u$ and $Y_d$, and the running of the CKM matrix is not sensitive to the flavor-diagonal parts in the beta functions for $Y_f$. Explicitly, one could insert the CKM matrix into Eq.~\eqref{eq:RGE-Y-quark}, and obtain the individual beta functions for the CKM mixing angles
\begin{eqnarray}
    \label{eq:CKM_t12} \dot{\theta}_{12} & = & -\frac{3}{2} \left(y^2_t + s y^2_t \right) c_{12} \left[ \left( s^2_{13} c^2_{23}-s^2_{23} \right) s_{12} +2s_{23} s_{13}
c_{12}c_{23}c_\delta\right] \; ,\\
    \label{eq:CKM_t23} \dot{\theta}_{23} & = & \phantom{-}\frac{3}{2} \left(y^2_t + s y^2_t \right) s_{23}c_{23} \; , \\
    \label{eq:CKM_t13} \dot{\theta}_{13} & = & \phantom{-}\frac{3}{2} \left( y^2_t + s y^2_t \right) s_{13}c_{13}c^2_{23} \; ,
\end{eqnarray}
as well as $\dot \delta \simeq 0$ at leading order.

Note that current experiments indicate that all the quark mixing angles are relatively small and the CKM matrix takes a nearly diagonal form. Thus, the beta function for $\theta_{12}$ [the right-hand side of Eq.~\eqref{eq:CKM_t12}] is strongly suppressed by sines of the mixing angles, implying that $\theta_{12}$ is stable against radiative corrections. The mixing angles $\theta_{23}$ and $\theta_{13}$ may receive visible RG corrections, and they increase with the energy scale. In fact, using Eqs.~\eqref{eq:CKM_t23} and \eqref{eq:CKM_t13}, it holds that $\theta_{23}$ and $\theta_{13}$ are related to each other by $\sin 2 \theta_{13} = C \tan \theta_{23}$, where $C$ is a constant.

\begin{figure*}[ht]
\begin{center}
\includegraphics[width=.75\textwidth]{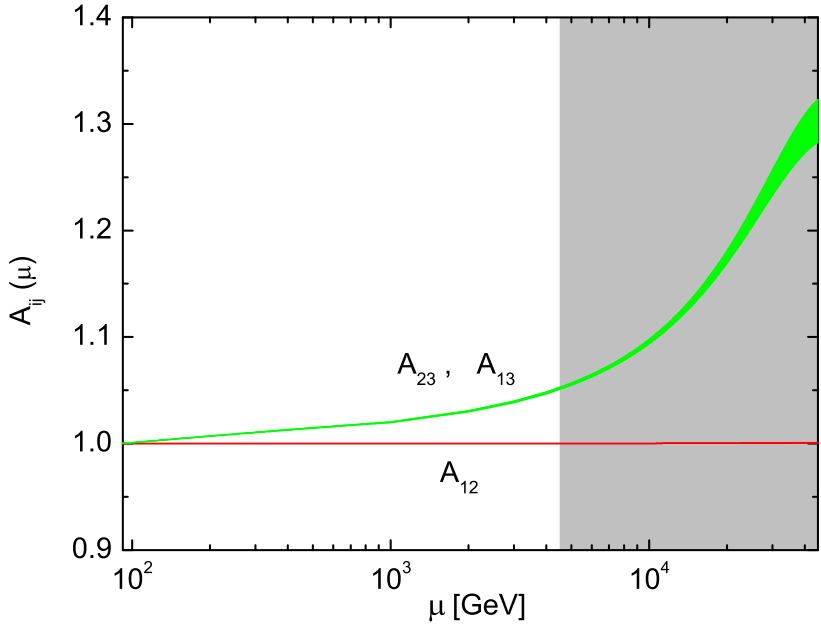}
\caption{\label{fig:theta} The RG evolution of the ratios $A_{ij}$ in the UED model with $R^{-1} = 1$~TeV. The curves of $A_{23}$ and $A_{13}$ overlap with each other. The gray-shaded area is ruled out by the vacuum stability criterion.}
\end{center}
\end{figure*}

In analogy with the mass ratios, we define the ratios $A_{ij} (\mu) \equiv \theta_{ij} (\mu)/\theta_{ij} (M_Z)$ characterizing the running behavior of the quark mixing angles. The evolution of the $A_{ij}$ is demonstrated in Fig.~\ref{fig:theta} for $R^{-1} = 1~{\rm TeV}$.
In agreement with our analytical results, $\theta_{12}$ is rather stable, whereas $\theta_{23}$ and $\theta_{13}$ could increase by about 5~\%.


\section{Summary and conclusions}\label{sec:Summary}

In this work, we have studied the RG running of physical parameters in the five-dimensional UED model. In particular, we have investigated the impact of the recent ATLAS\- and CMS Higgs mass bounds on the cutoff scale of the extra-dimensional model. These bounds come from the criterion of not reaching the Higgs vacuum instability limit below the cutoff scale. We have found that the five-dimensional UED model can be valid only at most up to the fifth KK level, significantly constraining the higher-dimensional description.

Using this new result, we have shown that it is generally not possible to achieve gauge coupling unification at the cutoff scale in the UED model. Furthermore, we have studied the RG running of the quark and lepton masses and mixing parameters in the UED model. We have found that, while the running at high-energy scales shows interesting features, these regions are excluded by the new bounds on the model. In particular, the regions of large power-law running is excluded.

Our results demonstrate that the LHC searches for the SM Higgs boson can have important consequences also for models of physics beyond the SM. As the bounds on the Higgs mass become even stronger, the global limit on the cutoff scale could be decreased sufficiently to allow for only three KK levels in the model. We emphasize that this is only an upper limit on the cutoff scale, and that new physics that changes the evolution of the Higgs self-coupling constant sufficiently to avoid the vacuum instability has to be introduced below this scale.

Although we have considered only the five-dimensional UED model, we expect that even stronger constraints can be derived for six-dimensional models, since a higher density of states gives rise to an even faster running in such models.


\begin{acknowledgments}

This work was supported by the Swedish Research Council (Vetenskapsr{\aa}det), contract
nos.~621-2008-4210 and 621-2011-3985 (T.O.), the ERC under the Starting Grant MANITOP and the DFG in the Transregio 27 ``Neutrinos and Beyond'' (H.Z.).

\end{acknowledgments}


\appendix

\section{One-loop beta functions in the UED model} \label{sec:appendix}

The running of the Higgs self-coupling constant is given by
\begin{eqnarray}\label{eq:RGE-lambda}
    16 \pi^2 \frac{{\rm d}\lambda}{{\rm d}\ln \mu} = \beta_\lambda + s \tilde \beta_\lambda \, .
\end{eqnarray}
Here, we have defined the scale parameter $s = \lfloor \mu/\mu_0 \rfloor$, where $\lfloor x \rfloor$ is the closest integer below $x$. The SM contribution reads~\cite{Cheng:1973nv}
\begin{eqnarray}
    \beta_\lambda  & = & 6 \lambda^2 - \lambda\left( 3 g^2_1 +9 g^2_2 \right) + \left( \frac{3}{2}g^4_1 + 3 g^2_1 g^2_2 + \frac{9}{2}g^4_2 \right) \nonumber \\
        && {} + 4 \lambda T - 8 {\rm tr} \left[ 3 \left( Y^\dagger_u Y_u \right)^2 + 3 \left( Y^\dagger_d Y_d \right)^2+ \left( Y^\dagger_\ell Y_\ell \right)^2 \right] \, .
\end{eqnarray}
where $T= {\rm tr} \left(3 Y^\dagger_u Y_u + 3 Y^\dagger_d Y_d + Y^\dagger_\ell Y_\ell \right)$. In addition, the extra-dimensional contributions are \cite{Bhattacharyya:2006ym}
\begin{eqnarray}
    \tilde \beta_\lambda & = & 6 \lambda^2 - \lambda\left( 3 g^2_1 + 9 g^2_2 \right) + \left( 2 g^4_1 + 4 g^2_1 g^2_2 + 6 g^4_2 \right) \nonumber \\
    & & {} + 8 \lambda T - 16 {\rm tr} \left[ 3 \left( Y^\dagger_u Y_u \right)^2 + 3 \left( Y^\dagger_d Y_d \right)^2 + \left( Y^\dagger_\ell Y_\ell \right)^2 \right] \, .
\end{eqnarray}

Next, the RGEs for the gauge couplings are given by
\begin{eqnarray}\label{eq:RGE-g}
    16\pi^2 \frac{{\rm d}g_i}{{\rm d}\ln \mu} = \left( b_i + s \tilde b_i \right) g^3_i \, ,
\end{eqnarray}
where $(b_1,b_2,b_3) = (41/6,-19/6,-7)$ and $(\tilde b_1,\tilde
b_2,\tilde b_3) = (27/2,7/6,-5/2)$ \cite{Bhattacharyya:2006ym}.

Finally, the one-loop RGEs for the Yukawa coupling matrices $Y_f$ ($f=u,d,\ell$) and the neutrino mass operator $\kappa$ can be expressed in a general form as
\begin{eqnarray}\label{eq:RGE-Y-quark}
    16 \pi^2 \frac{{\rm d}Y_f}{{\rm d}\ln \mu} & = & \beta_f + \tilde \beta_f = \beta_f + s \tilde \alpha_f Y_f + s Y_f \tilde N_f \, ,\\
    \label{eq:RGE-kappa} 16 \pi^2 \frac{{\rm d}\kappa}{{\rm d}\ln \mu} & = & \beta_\kappa + \tilde \beta_\kappa =  \beta_\kappa + s \tilde \alpha_\kappa \kappa + s \kappa \tilde N_\kappa + s \tilde N^T_\kappa \kappa \, ,
\end{eqnarray}
where the SM beta functions are~\cite{Babu:1993qv,Chankowski:1993tx,Antusch:2001vn}
\begin{eqnarray}
    \beta_u  & = & Y_u \left(\frac{3}{2} Y^\dagger_u Y_u - \frac{3}{2} Y^\dagger_d Y_d - \frac{17}{12} g^2_1 - \frac{9}{4} g^2_2 - 8g^2_3 + T \right) \, ,\\
\beta_d & = & Y_d \left( -\frac{3}{2}  Y^\dagger_u Y_u + \frac{3}{2} Y^\dagger_d Y_d - \frac{5}{12} g^2_1 - \frac{9}{4} g^2_2 - 8g^2_3 + T \right)\, ,\\
    \label{eq:beta-l} \beta_\ell & = & Y_\ell \left( \frac{3}{2}  Y^\dagger_\ell Y_\ell - \frac{15}{4} g^2_1 - \frac{9}{4}g^2_2+T\right) \, , \\
    \label{eq:beta-kappa-sm} \beta_\kappa & = & -\frac{3}{2} \kappa \left( Y^\dagger_\ell Y_\ell \right)-\frac{3}{2} \left( Y^\dagger_\ell Y_\ell \right)^T \kappa + \left(\lambda - 3 g^2_2 + 2 T \right) \kappa \, .
\end{eqnarray}
The contributions from the KK excitations are given by \cite{Cornell:2010sz,Blennow:2011mp}
\begin{eqnarray}
    \tilde N_u  & = & \phantom{-}\frac{3}{2} Y^\dagger_u Y_u - \frac{3}{2} Y^\dagger_d Y_d \, ,\\
    \tilde N_d  & = & -\frac{3}{2} Y^\dagger_u Y_u + \frac{3}{2} Y^\dagger_d Y_d \, ,\\
    \tilde N_\ell & = & \phantom{-}\frac{3}{2} Y^\dagger_\ell Y_\ell \, , \\
    \tilde N_\kappa & = & -\frac{3}{2} Y^\dagger_\ell Y_\ell \, ,
\end{eqnarray}
and
\begin{eqnarray}\label{eq:alphas}
    \tilde \alpha_u & = & - \frac{101}{72} g^2_1 - \frac{15}{8} g^2_2 - \frac{28}{3}g^2_3 + 2T \, , \\
    \tilde\alpha_d & = & -\frac{17}{72} g^2_1 - \frac{15}{8} g^2_2 - \frac{28}{3} g^2_3 + 2T \, , \\
    \tilde\alpha_\ell & = & - \frac{33}{8} g^2_1 - \frac{15}{8} g^2_2 + 2 T \, , \\
    \tilde\alpha_\kappa & = &  -\frac{1}{4} g^2_1 - \frac{11}{4} g^2_2 + 4 T + \lambda \, .
\end{eqnarray}


\end{document}